\documentclass[aps,prc,twocolumn,superscriptaddress]{revtex4}
\usepackage{amssymb}
\usepackage{amsmath}
\usepackage{CJK}
\usepackage{graphicx}
\usepackage{xcolor}
\newcommand{\Tr}{{\mathrm{Tr}}}
\usepackage{array,multirow}
 \usepackage{color}
 \usepackage[normalem]{ulem}
\begin{document}
\title{Entanglement and seniority}

\author{J. Kov\'acs}
\affiliation{Institute for Nuclear Research, Debrecen, Hungary}
\author{A. T. Kruppa}
\affiliation{Institute for Nuclear Research, Debrecen, Hungary}
\affiliation{Department of Theoretical Physics, Budapest University of Technology and Economics, H-1111 Budapest, Hungary}
\author{P. Salamon}
\affiliation{Institute for Nuclear Research, Debrecen, Hungary}
\author{\"O. Legeza}
\email{legeza.ors@wigner.hu}
\affiliation{Wigner Research Centre for Physics, H-1525 Budapest, Hungary}
\affiliation{Fachbereich Physik, Philipps-Universit\"at Marburg, 35032 Marburg, Germany}
\affiliation{Institute for Advanced Study, Technical University of Munich, Lichtenbergstrasse 2a, 85748 Garching, Germany}
\author{G. Zar\'and}
\affiliation{Department of Theoretical Physics,  Budapest University of Technology and Economics, H-1111 Budapest, Hungary}
\affiliation{MTA-BME Quantum Dynamics and Correlations Research Group, E\"otv\"os Lor\'and Research Network (ELKH),  Budapest University of Technology and Economics}

\begin{abstract}
We study mode-entanglement in the seniority model,   derive analytic formulas 
for the one-body reduced density matrix  of states 
with seniority $\nu = 0,\;1,\;2$, and $\nu=3$, and also  determine the particle number dependence  
of the one-body reduced density matrix for arbitrary seniority.
We carry out numerical calculations for the lightest calcium isotopes and for  $^{94}{\rm Ru}$ nucleus,  
and investigate the structure of their ground and low energy yrast states.
We investigate the fulfillment of several predictions of the seniority model regarding the behavior of one-mode entropies,
which we compare with the results of   {configuration interaction}  (CI) and  density matrix renormalization group  {(DMRG)} computations.
For $^{94}{\rm Ru}$, the seniority model accounts  for the $0g_{9/2}$ mode entropies, but seniority mixing is
important for certain  yrast states. Interaction induced quantum fluctuations decrease the occupation of the 
$0f_{5/2}$, $1p_{3/2}$ and $1p_{1/2}$  shells, and amount in finite mode entropies on these shells, too, 
clearly outside the scope of the simple $(0g_{9/2})^4$ seniority model. The $0f_{7/2}$ shell based seniority model is more 
accurate for  the light ${\rm Ca}$ isotopes, but seniority mixing is substantial for some $^{44}{\rm Ca}$ yrast states, too. 
\end{abstract}

\maketitle

\section{Introduction}

In recent decades, considerable effort has been devoted to entanglement in many areas of physics. 
Besides  investigations motivated by the application of entanglement as a resource \cite{Horodecki}, 
more and more attention is devoted  to the structure and role of entanglement in many-body problems~\cite{Amico,Eisert,Laflorencie}. 
While entanglement is an early concept of quantum theory and  has been the subject of intense  investigations \cite{chen}, 
in systems of indistinguishable particles, such studies do not have a long history. In the latter case, 
the definition of subsystems is a more subtle question,  since decomposition based on 
the tensor product of Hilbert spaces does not lead to  physically meaningful subsystems.
Different approaches are introduced to overcome this problem: the mode entanglement method   
\cite{zan02,git02,leg03,shi03,leg04,fri13,gig15,gig21}, the algebraic approach based on the correlation between observables
\cite{ris06,ban07,bal13,ben12,ben14,bar15,ben20,szal20}, descriptions relying on quantum correlations of 
particles \cite{pas01,sch01,eck02,Ghirardi} and concepts generalized to quasi-particles \cite{gig15,tul18}.

In this paper, we follow the formalism based on algebraic 
partitions of bounded operators generated by the fermionic creation and annihilation operators,
and we apply the entanglement measure called one-mode entropy \cite{leg03,gig15}.
The one-mode entropy being, however, basis dependent \cite{leg03,gig15},
we also utilize the  notion of  basis independent  one-body entanglement entropy, introduced  in Refs.~\cite{kru15,gig15,kru21} 
to characterize entanglement.

Entanglement and correlations are somewhat related concepts.
 Correlations  play an unavoidable role in non-perturbative many-body problems \cite{Tew,tic11,sza15,sza17}
and, -- similar to entanglement, --  characterize the connections or independence of certain quantities or subsystems. 
There is therefore  a natural demand to investigate concrete correlated quantum mechanical models from the viewpoint of 
entanglement. This paper aims to contribute to  this line of investigations by the study of 
the seniority model~\cite{Talmi} in nuclear physics, and by comparing its predictions for entanglement 
with detailed model calculations.

Entanglement investigations are often performed  in atomic 
physics \cite{qva20,deh11}, quantum chemistry \cite{sza15},  and condensed matter physics~\cite{Laflorencie}
to date.  Although investigations of the Lipkin-model \cite{tul19,fab21} and fermionic superconducting systems \cite{tul18}, 
do have some relevance for nuclear physics, exploration of this research area in the context 
of nuclear physics is, however,  still in its initial stage. Apart from studies carried out in 
the traditional nuclear shell model framework \cite{kva14,kva16,leg15,gor18,gor19,kruppa21} and an investigation performed 
in an \emph{ab initio} no-core shell model \cite{sav}, we are not aware { of other works persuing} this research direction. 

In this paper, we study few nucleon states within the seniority model. This investigation can be considered as an
extension of our previous study  on the entanglement of angular-momentum coupled 
two-nucleon states~\cite{kruppa21}  to relatively simple many-nucleon states. 
 {One-mode} entanglement and two-orbital correlations in wave 
functions restricted to seniority zero electron pair states have also been studied 
in Ref.~\cite{Bog16} in a quantum chemistry context.  There, however, a  general framework is used, 
without exploiting the conservation of total angular momentum, an almost mandatory constraint
in nuclear physics.

The physical motivation behind  the seniority model is the pairing phenomenon~\cite{Talmi}.
 The seniority model can be considered as a first step to describe 
 short-range correlations in nuclei, but can also be used as an essential building block for more sophisticated 
models such as the generalized seniority model (see, e.g.,Refs.~\cite{tal71,shlomo72,luo11}).
The seniority model is a $jj$-coupling classification of nucleons residing on a single-$j$ shell, with 
the seniority quantum number $\nu$  naturally interpreted as the  number of unpaired particles. 

From a mathematical viewpoint, the seniority model is 
an exactly solvable model describing $n$ particles interacting via  seniority conserving  interaction,  
and possessing a dynamically broken  SU(2) quasi-spin symmetry~\cite{ros03}. 
Even though realistic  interactions with  seniority mixing  do not possess this dynamical symmetry, 
the classification of states based on seniority quantum numbers often proves very useful 
for interpretation.  Although the seniority model and its generalization to multi-$j$ shells
 appeared  early in nuclear physics, they provide a valid and quite 
 accurate description  for many nuclei, and a seniority quantum number-based classification 
 is still quite often applied~ \cite{Val21,sam20,mah19,sto19,mil19}. 
Therefore, proceeding along these lines,  
we compute in this work the mode entropy within the seniority model, and compare its
predictions with the results of many-body  density matrix renormalization group (DMRG)\cite{White1992} 
and configuration interaction (CI)  computations, performed  for the lightest calcium isotopes and 
for $^{94}{\rm Ru}$. Both belong to the family of  semi-magic  nuclei. 
In the  case of Ca, the proton shell is closed and relevant neutrons dominantly occupy the outer $0f_{7/2}$ shell,   
while  in $^{94}{\rm Ru}$ the neutron shell is closed, and  the relevant $0g_{9/2}$  (open) proton shell  hosts four protons.
Low-lying excited states can be interpreted in both cases as seniority $\nu=2$ pair breaking states
with an admixture of seniority  $\nu=4$ states.

It is important to mention that there are two particular seniority $\nu=4$ states \cite{esc06,zam07,isa08,qi11,qia18} 
with total angular momentum four and six, which can be shown to be 
  eigenstates of \emph{any} two-body interaction restricted to the $(9/2)^4$ configuration. 
These states play an essential role in the explanation of the seniority isomerism and electromagnetic 
transitions \cite{qi17,isa14,isa11,mor18} and, as we shall see, 
they are also important from the perspective of entanglement.

The paper is organized as follows. We review the concept of  mode entanglement  
 in Section~\ref{ent1}.  The basic concepts of the seniority model are  
presented in Section~\ref{chapsen}. States with seniority $\nu=0$, $\nu=1$, and $\nu=2$  are analyzed from the perspective 
of  mode-entanglement in Section~\ref{single}. Section~\ref{anres} discusses our analytical results, while 
numerical results are presented in Section~\ref{numres}, where the lightest calcium isotops and
 $^{94}{\rm Ru}$ nuclei are investigated by means of the  configuration interaction (CI) method and 
a  nuclear shell version of the density matrix renormalization group (DMRG) method~\cite{Dukelsky2002,leg15}, 
using realistic effective interactions. Results obtained by the CI and DMRG methods
are compared with the predictions of the seniority (SEN) model. 
Finally, Section~\ref{sum} concludes and summarizes the results.

\section{Mode entanglement}\label{ent1}

In this work, we apply the second quantized formalism to determine 
mode-entanglement in a many-body framework.
The primary objects for a fermionic system of $d$ modes are 
the creation and destruction  operators, $c_i^\dagger$ and $c_i$, $i \in \{1,2,\ldots,d\}$, 
  satisfying  canonical anti-commutation relations.
The single-particle (sp) states $c^\dagger_i\vert 0\rangle$ span the single particle Hilbert space of dimension $d$, 
with $\vert 0\rangle$ referring to the vacuum.
Considering a bipartition of the $d$ modes to subsets $A$ and $B$, we get two 
algebras, ${\cal A}_A$, and ${\cal A}_B$,  spanned  by operators composed from the sets $A$ and $B$, respectively.  
A purely  fermionic state is separable with respect to  this bipartition if and only if it is of 
the form 
\begin{equation}\label{sep}
{\cal P}(\{c_{i\in A}^\dagger\}) \, {\cal Q}(\{c_{j\in B}^\dagger\})\vert 0\rangle,
\end{equation}
with ${\cal P}$ and ${\cal Q}$ denoting polynomials of the creation operators. In this work, we 
restrict ourself to the  case, where $A$ refers to a \emph{single} mode,   $k$, while  $B$ contains all other $i\ne k$ modes, and 
the whole system is in a pure state, $\Psi$.

To characterize this type of bipartition, one introduces the one-mode reduced density matrix, 
\begin{equation}\label{onemden}
 \rho_{k}\equiv \left(\begin{array}{cc}
\langle c_k^\dagger c_k\rangle&0\\
0& 1-\langle c_k^\dagger c_k\rangle
\end{array}\right),
\end{equation}
with  $\langle c_k^\dagger c_k\rangle=\langle\Psi\vert c_k^\dagger c_k\vert\Psi\rangle$. 
The entanglement measure of this simple  bipartition is the one-mode entropy, 
\begin{equation}\label{onemdef}
{\cal S}_k\equiv -\Tr( \rho_{k}\,{\ln}( \rho_{k}))=h(\langle c_k^\dagger c_k\rangle),
\end{equation}
with the function $h$  defined as 
$$
h(x)\equiv -x\, {\ln}(x)-(1-x)\,{\ln}(1-x)\;.
$$
The  total correlation introduced as 
\begin{equation}
{\cal S}_c=\sum_{k=1}^d {\cal S}(\rho_{k}).
\end{equation}
yields then a global characterization of the entanglement of the state $\Psi$~\cite{leg04}, which 
 depends, however,  on the choice of  sp basis, $\{c\}$.
A basis independent  one-body entanglement entropy can then be defined as \cite{kru15,gig15,kru21} 
\begin{equation}
{\cal S}_{\rm 1B}  =   {\min_{\{c\}}} \;{\cal S}_{c}\;.
\end{equation}
By the measure ${\cal S}_{1B}$, all fermion states described by a single Slater determinant 
are classified as  non-entangled, while other pure 
states are entangled~\cite{sch01,eck02,gig15}.

The quantity $S_{\rm 1B}$ is closely related to the eigenvalues of the one-body reduced density matrix (1B-RDM),  
 defined as
\begin{equation}
 \rho_{ij}=\langle\Psi\vert c_j^\dagger c_i\vert\Psi\rangle,
\end{equation}
and normalized as ${\rm Tr}\;\rho=n$, with $n$  the particle number.
The basis where ${\cal S}_{\{c\}}$ reaches its minimum (\emph{i.e.}, ${\cal S}_{\{c\}}=S_{\rm 1B}$) corresponds to the 
so-called natural orbitals, and there  the 1B-RDM is diagonal,   $ \rho={\rm diag}\{n_1,n_2,\ldots,n_d\}$~\cite{gig15,kru21}.
There ${\cal S}_{\rm 1B}$ reads
\begin{equation}\label{onespent}
{\cal S}_{\rm 1B}=\sum_{i=1}^d h(n_i),
\end{equation}
where the $n_i$ denote the occupation numbers of the natural orbitals.

\section{Seniority model: basic definitions}\label{chapsen}

In the seniority model, we assume that $n$ nucleons occupy a single $j$ shell, corresponding to a shell configuration  $j^n$.
For the sp states, we use the notation $\vert a,m\rangle=\vert n_al_aj_am\rangle$. 
In the seniority model, nucleons reside on a single multiplet, $a$. We can therefore 
suppress the label $a$ and  use the compact notations, $a^\dagger_{am}\to a^\dagger_m$ and $a_{am}\to a_m$
for the corresponding creation and annihilation operators. 

The pair creation and annihilation  operators, $S_\pm$,  defined as 
\begin{equation}
S_+=\sum_{m>0}(-1)^{j-m} a^\dagger_{m} a^\dagger_{-m}\phantom{nn}\textrm{and} \phantom{nn} S_- = S_+^{\,\dagger}
\end{equation}
create/destroy a zero angular momentum pair of particles.
Together with
\begin{equation}
S_0
=\frac{1}{2}\Big(\sum_m a_{m}^\dagger a_{m}-{ \frac {2j+1}2}\Big)\;,
\end{equation}
the $S_\pm$ obey  standard  SU(2) angular momentum commutation relations, and the  operators $S_x = (S_+ + S_-)/2$, $S_y = (S_+ - S_-)/2i$, 
and $S_z = S_0$ form  components of the so-called  quasi-spin operator, $\bf S$. 

In the seniority model, a state  of angular momentum $J$ and projection $J_z = M$ is denoted by  $\Psi_{JM}(j^n\nu\, \,\eta)$, where   
$\,\eta$ stands for an angular momentum multiplicity label. 
If the state is multiplicity-free, the index $\,\eta$ can (and will) be suppressed. 
In the seniority model, we start from unpaired  $\nu$-particle states, which do not contain paired particles,
$S_-\Psi_{JM}(j^\nu\,\nu\,\eta)=0$,  and  are quasi-spin eigenstates of spin  $S = (j + \frac 1 2 -\nu)/2$ 
and $S_z= - S$. The quantum number $\nu$  defines the \emph{seniority} of this and all descendent 
states.  From the properties of the quasi-spin operators it follows that the $n$-particle state  
\begin{equation}\label{gennu2}
\psi_{JM}(j^n\,\nu\,\eta)={\cal N}_{n,\nu} \, S_+^{(n-\nu)/2}\, \psi_{JM}(j^\nu\,\nu\,\eta)
\end{equation}
has the same quasi-spin as $\psi_{JM}(j^\nu\,\nu\,\eta)$, but is an eigenstate of  
$S_z$ with eigenvalue   $-(j+\frac 1 2 -n)/2$. The prefactor  ${\cal N}_{n,\nu}$ here just 
 ensures proper normalization.  Clearly, instead of the quantum numbers $n$ and $\nu$, 
 we can thus use the values of the quasi-spin $S$ and its third component $S_z$ 
as quantum numbers,  and denote the state  $\Psi_{JM}(j^n\nu\,\eta)$ as  $\Psi_{JMSS_z}(j\,\eta)$. 

If all  nucleons are paired, the $n$-particle seniority-zero wave function reads
\begin{equation}
\Psi_{00}(j^n0)=
{\cal N}_{n,0}\, S_+^{\,n/2}\vert 0\rangle.
\end{equation}
For an odd number of particles, there is only one unpaired particle {in the ground state},
 and the 
corresponding seniority-one eigenfunction is
\begin{equation}
\Psi_{jM}(j^n1)
={\cal N}_{n,1} \,S_+^{(n-1)/2}a^+_{M}\vert 0\rangle.
\end{equation}
Seniority-two states have the form
\begin{eqnarray}
&&\Psi_{JM}(j^n2)=
{\cal N}_{n,2}\, S_+^{(n-2)/2}
\\
&&\phantom{nnnn}\Big(\sum _mC_{ j,m\, j,M-m}^{ J,M} \,a^\dagger_{m}a^\dagger_{M-m}\Big)\vert 0\rangle,
\nonumber
\end{eqnarray}
where $J=2j-1,\, 2j-3, \dots{,2}$.
In the cases of $\nu=0$, $\nu=1$ and $\nu=2$ the states are multiplicity free. The form of higher seniority states 
is much more complicated, and 
analytical forms are only known for $\nu=3,4$ and $5$~\cite{kan80,che19}. 

\section{One-body reduced density matrix}\label{single}

To determine the 1B-RDM,  we consider first a general, model-independent state,  $\Psi_{JM}$, with  angular momentum,
$J$, and $J_z = M$.   The 1B-RDM is then given by 
\begin{equation}\label{oprdmgen}
\rho^{\;a^\prime a}_{m^\prime m}
(JM)=\langle \Psi_{JM}\vert a_{a'm'}^\dagger a_{am}\vert \Psi_{JM}\rangle\;.
\end{equation}
Conservation of the $z$ component of the angular momentum implies that all
$m\ne m^\prime$ off-diagonal elements of $\rho^{a'a}_{m'm}(JM)$ vanish. In the special case, $J=0$, moreover, 
simple group theoretical arguments imply  $j_a = j_{a'}$, and that all diagonal matrix elements are equal
\begin{eqnarray}\label{j0res}
\rho^{\;a^\prime a }_{m^\prime m }
(00)={\delta_{m^\prime m}\delta_{j_{a^\prime} j_{a}}}  \, \hat \rho_{a^\prime a} (00)\;.
\end{eqnarray}
This immediately  implies that in any $J=0$ state  the one-mode entropies do not depend on the 
magnetic quantum number $m$ of the sp modes, as also observed numerically in Ref.~\cite{leg15} and 
proved in \cite{kruppa21} for two-body problems.

To compute  the matrix elements of the operators $a_{a'm'}^\dagger a_{am}$ for $J\ne 0$,
we express these in terms of spherical tensor operators. The
operators $a_{am}^\dagger$ and $\tilde a_{am}\equiv (-1)^{j_a+m}a_{a,-m}$ are both  tensor operators of 
rank (spin)  $j_a$. We  can define from these
spherical tensor operators of rank $K$ using the usual SU(2) addition rules as 
\begin{eqnarray}
\label{tensor}
&&\left[a_{a^\prime}^\dagger\otimes \tilde a_{a}\right]^{K}_{\,k} =
\sum_{m} C_{ j_{a^\prime}, m,\,j_{a},\,k-m}^{ K,\,k}\, a_{a^\prime,m}^\dagger \tilde a_{a,k-m}\;,
\nonumber
\end{eqnarray}
and express  the 1B-RDM as 
\begin{eqnarray}\label{calcv2a}
&&
\rho^{\;a^\prime a}_{m^\prime m}(JM)= \delta_{m^\prime,m} \,(-1)^{j_a-m}
\\
&&\phantom{nnn}\sum_{K } C_{ j_{a^\prime}, m,\, j_a,-m}^{ K, 0}
\langle \Psi_{JM}\vert \big[a_{a^\prime }^\dagger\otimes \tilde a_{a}\big]^{K}_{\,0}\vert \Psi_{JM}\rangle.
\nonumber 
\end{eqnarray}
Clearly, for $J=0$ only the $K=0$ term remains, and  the expression simplifies to Eq.~\eqref{j0res}.

Our goal is to calculate the  1B-RDM, $\rho_{m^\prime,m}(JM,j^n\nu\,\eta)$,   associated with 
 the state $\Psi_{JM}(j^n\nu\,\eta)$ within  the seniority model.  We shall do that using the  
 quasi-spin formalism~\cite{Lawson}, where we   introduce a quasi-spin tensor of rank 
 $\frac{1}{2}$ as
\begin{equation}
R^{\frac{1}{2}}_{\mu;m}
=\left\{
\begin{array}{ll}
a^\dagger_{m}&\quad{\rm if\ \ }\mu=1/2\;,\\
-\tilde a_{m}&\quad{\rm if\ \ }\mu=-1/2\; .
\end{array}\right.
\end{equation}
From these  we define quasi-spin tensors of rank $K=0$ and $K=1$ as \cite{Lawson}
\begin{equation}\label{defrkq}
R^{K}_{\;\kappa;\; m^\prime, m }=\sum_{\mu,\mu^\prime}
C_{\; {\textstyle  \frac{1}{2}},\,\mu^\prime,\,  {\textstyle  \frac{1}{2}},\,\mu}^{K, \kappa } \;
R^{\frac{1}{2}}_{\mu^\prime ;m^\prime}R^{\frac{1}{2}}_{\mu;m}\;.
\end{equation}
The nucleon number dependence of the matrix element of the operator $a_{m^\prime}^\dagger a_{m\phantom{^\prime}}$ can be determined by using 
 the identity
\begin{equation}\label{aop}
a^\dagger_{m^\prime}a_{m}=
\frac{(-1)^{j+m}}{\sqrt{2}}\left( R^{0}_{\;0\,;\, m^\prime,-m}+R^{1}_{\;0\,;\,m^\prime,-m}
\right)\;,
\end{equation}
and by applying  the Wigner-Eckart theorem in the quasi-spin space, yielding
\begin{eqnarray}\label{wigner}
&&\langle \Psi_{JMSS_z}(j\,\eta)\vert a^\dagger_{m^\prime}a_{m}\vert\Psi_{JMSS_z}(j\,\eta)\rangle=
\delta_{m^\prime\, m}
\frac{(-1)^{j+m}}{\sqrt{2}}\nonumber\\
&&\phantom{nn}\Big \{
C_{ S,S_z,0,0}^{ S,S_z} \langle \Psi_{JMS}(j\,\eta)\nmid\nmid R^{0}_{m^\prime,-m}\nmid\nmid\Psi_{JMS}(j\,\eta)\rangle\nonumber\\
&& \phantom{nn} +C_{S,S_z,1,0}^{ S,S_z} \langle \Psi_{JMS}(j\,\eta)\nmid\nmid R^{1}_{m^\prime,-m}\nmid\nmid\Psi_{JMS}(j\,\eta)\rangle
\Big\},
\nonumber
\end{eqnarray}
where $\nmid\nmid $ indicates  the reduced matrix element in  quasi-spin space.
 By using the
explicit values of the Clebsch-Gordan coefficients 
and the relations between $S$, $S_z$, $n$, and $\nu$, 
we obtain 
\begin{eqnarray}\label{freduct}
&&\rho_{m^\prime m }(JM,j^n\nu\,\eta)=
\\
&&{\textstyle \frac{(-1)^{j+m}}{\sqrt{2}}}\big[\langle \Psi_{JM}(j^\nu\nu\,\eta)\vert R^{(0)}_{\;0;\; m^\prime,-m} \vert\Psi_{JM}(j^\nu\nu\,\eta)\rangle
\nonumber\\
&&+{\textstyle \frac{2(j-n) +1 }{2(j-\nu) +1}}
\langle\Psi_{JM}(j^\nu\nu\,\eta)\vert R^{(1)}_{\;0;\; m^\prime,-m}\vert\Psi_{JM}(j^\nu\nu\,\eta)\rangle\big]\;.
\nonumber
\end{eqnarray}
This formula allows us to relate the  1B-RDM of a system of $n$ particles to that  of a system of $\nu$
particles, both in states with seniority $\nu$.

We now proceed and give the analytical expression for the 1B-RDM for wave functions with seniority $\nu=0, \,1$, and $\nu=2$. 
Details of the derivations are  to  Appendix~\ref{appendix}.  In the simplest case, $\nu=0$,  the total angular momentum vanishes,
$J=0$, and therefore $\rho_{m^\prime m}$ is proportional to the unit matrix (see also Eq.~\eqref{j0res}),
\begin{equation}
\label{nu0om}
\rho_{m^\prime m }(00,j^n0)=\delta_{m^\prime,m}\;\frac{n}{2j+1}\;.
\end{equation}

For $\nu=1$, only the sp orbitals $m=\pm M$ behave differently from the rest.
In this case, the $m=M$ orbital is occupied, while $m=-M$ is empty, and the remaining 
$n-1$ particles reside on the other $2j-1$ orbitals with uniform probability, 
\begin{eqnarray}\label{nu1om}
\rho_{m^\prime m}(jM,j^n1)=
\delta_{m^\prime\,m}\left\{\begin{array}{ll}
1&\quad {\rm if }\quad m=M\;,\\
0&\quad {\rm if }\quad m=-M\;,\\
\frac{n-1}{2j-1}&\quad {\rm if} \quad m\neq\vert M\vert\;.
\end{array}\right.
\end{eqnarray}

\begin{widetext}
For seniority-two  wave functions, the 1B-RDM has the form
\begin{eqnarray}\label{nu2om}
\rho_{m^\prime m}(JM,j^n2)&=&\delta_{m^\prime\,m}\Big\{
\left(C_{ j,m,j,M-m}^{J,M}\right)^2
-\left(C_{ j,-m,j,M+m}^{J,M}\right)^2 +{\textstyle \frac 1 2}
\nonumber\\
&+& {\textstyle \frac{2(j-n)+1}{2j-3}}\Big[
\left(C_{ j,m,j,M-m}^{ J,M}\right)^2 
+\left(C_{ j,-m,j,M+m}^{J,M}\right)^2
-{\textstyle \frac 1 2} \Big]\Big\},
\end{eqnarray}
\end{widetext}
where the  total angular momentum is even and ranges  from $J=2$ to $J=2j-1$.
The complicated analytical form of the 1B-RDM is given in the Appendix for seniority-three states.

\section{Analytical results}
\label{anres}
We now discuss the consequences of the previous results from the perspectives of  the mode- and one-body 
entropies.
Having constructed  the 1B-RDM, we obtain  the mode entropies  by Eq.~\eqref{onemdef} as
\begin{equation}
{\cal S}^{\,a}_m(JM)=h\left(\rho^{\,a\, a}_{mm}(JM)\right).
\end{equation}
A mode $m$ of a multiplet $a$ is therefore non-entangled if and only if $\rho^{\,aa}_{mm}(JM)=0$ or  
$\rho^{\,aa}_{mm}(JM)=1$, i.e., if it is completely empty (the mode is not in the wave function) or completely occupied (each term of the CI expansion contains the mode). 
{From Eq.~(\ref{calcv2a}) it follows that for any state of  total angular momentum $J$ and projection  $M=0$, 
the modes $m$ and $-m$ of a multiplet $a$ have identical mode entropies,  ${\cal S}^a_m(J\,0)={\cal S}^a_{-m}(J\,0)$.
A further consequence of  (\ref{calcv2a}) is that  the one-mode entropies of the state $\Psi_{J\,M}$ are related
to those of the state $\Psi_{J\,-M}$ as  ${\cal S}^a_m(J\,{-M})={\cal S}^a_{-m}(J\,M)$.

Also,  states related by particle-hole transformation have identical  {one-body entanglement} entropies. Particle-hole conjugation,  is  represented  by a unitary operator, $\Gamma$,   having the properties~\cite{Lawson}
\begin{equation}\label{prop1}
\Gamma \,\Psi_{JM}(j^n\nu)=(-1)^{\frac{n-\nu}{2}}\Psi_{JM}(j^{2j+1-n}\nu),
\end{equation}
and 
\begin{equation}\label{prop2}
\Gamma^\dagger a_m^\dagger \Gamma=-\tilde a_{m}\;,  \quad  \Gamma^\dagger \tilde a_m \Gamma= a^\dagger_{m}\;.
\end{equation}
From these relations one immediately obtains the relation 
 $\rho_{\tilde m\,\tilde m}(JM,j^{2j+1-n}\nu)=1-\rho_{m\,m}(JM,j^{n}\nu)$, with $\tilde  m = -m$ referring to the time-reversed orbital. 
Since  $h(1-x)=h(x)$,  this immediately implies  
${\cal S}_{-m}(JM,j^{2j+1-n}\nu)={\cal S}_{m}(JM,j^n\nu)$, and the relation
\begin{equation}
{\cal S}_{\rm 1B}(JM,j^{2j+1-n}\nu)={\cal S}_{\rm 1B}(JM,j^n\nu)\,,
\end{equation}
where  we introduced the notations
${\cal S}_m(JM,j^n\nu\,\eta)$ and ${\cal S}_{\rm 1B}(JM,j^n\nu\,\eta)$ for mode- and one-body  entropies within the seniority model.
Thus $n$-particle and $n$-hole states  have the same one-body entanglement entropy.

For seniority $\nu=0$ wave functions the 1B-RDM is diagonal (see Eq.~\eqref{nu0om}),
and  the one mode entropy is  mode independent, ${\cal S}_m  \to  h(n/(2j+1))$. The one-body  entropy is 
therefore
\begin{equation}
{\cal S}_{\rm 1B}(00,j^n0)=(2j +1)\, h\left(\textstyle{\frac{n}{2j+1}}\right).
\end{equation}
For seniority-zero states the largest ${\cal S}_{1B}$ therefore corresponds to a half-filled shell, $n=j + \frac 1 2$. 

One may ask, which state is having the largest one-body entropy from all possible wave functions of a configuration $j^n$? 
It has been  shown that the total correlation has its minimum value in the natural basis \cite{gig15,kru21}, and the upper limit of 
the one-body entanglement  entropy is reached when the 1B-RDM is proportional to the unit matrix, $\rho^\textrm{max}_{m^\prime\,m}(JM,j^n\nu\,\eta)=\frac{n}{2j+1}\,\delta_{m^\prime\,m}$~\cite{kruppa21}. 
This criterion is satisfied by seniority-zero wave functions  as well as by any  $J=0$ wave function, 
which  are therefore \emph{  maximally entangled }  states with respect to  one-body entanglement.

Seniority  $\nu=1$ states must contain a broken pair, and are  less entangled. Indeed, in the state $\Psi_{jM}(j^n1)$,
 the modes with  $m=\pm M$  
are occupied and empty with probability one (see  Eq.~(\ref{nu1om})), and are therefore non-entangled. 
All other $m\ne \pm M$  modes, -- hosting the remaining $(n-1)/2$ pairs of particles, -- have 
  one-mode entropies  $h((n-1)/(2 j-1))$. 
 We thus obtain the following one-body entropy for seniority $\nu=1$ states,  
\begin{equation}
{\cal S}_{1B}(jM,j^n1)=(2j-1)\;h\Big(\textstyle{\frac{n-1}{2j-1}}\Big)\;.
\end{equation}
Thus for $\nu=1$, too,  the largest one-body entanglement entropy corresponds to a half-filled shell, just  as for states with 
$\nu=0$. Notice that ${\cal S}_{1B}(jM,j^n1)$ is independent of $M$.

The one-body entropy for seniority $\nu=2$ states is more complicated, in general. 
 An interesting situation arises, however, 
if $j+\frac 1 2$ is an even number, $M=0$, and we consider a half-filled shell, 
$n=j+\frac 1 2$. Then each mode is maximally entangled, ${\cal S}_{m}(J0,j^{j+ 1/ 2 } \,2)=\ln 2$ since 
$\rho_{m\,m}(J\,0,j^{j+ 1/ 2 } \,2)=1/2$, and ${\cal S}_{\rm 1B}(J\,0,j^{j+ 1/ 2 }  \,2)=(2j+1)\,\ln 2 $.
We thus conclude that maximally one-body  entangled states exist  even with broken pairs, 
$\nu\ne 0$ and $J\ne0$.

Typically, all modes are entangled to some degree in seniority two states. 
Non-entangled modes  can be observed, however, when the angular momentum has its maximal value, $J=2j-1$,
and $M=2j-1$ or $M=2j-2$. In the case $M=2j-1$, the modes $m=j$ and $m=j-1$ and their time-reversed pairs, $m=-j$ and $m=1-j$, are non-entangled,
 while  for  $M=2j-2$  the modes $m=j$ and $m=j-2$ and their pairs,  $m=-j$ and $m=2-j$ become non-entangled. These statements 
 can be directly verified by using the expression \eqref{nu2om}, 
but a physical explanation can also be given. In the two-particle  wave function $\Psi_{2j-1,2j-1}(j^2\,2)$, e.g., 
the states $m=j$ and $m=j-1$ are occupied, while all other states
are empty. 
Applying the operator $S_+^{(n-2)/2}$ on this state leaves the states  $m=j$ and $m=j-1$  occupied and 
their time reversed pairs,  $m=- j$ and $m=-(j-1)$ empty, since $S_+$ can populate  
time reversed pairs $\{\pm m\}$ only 
simultaneously~\footnote{In other words, the wave function $\Psi_{2j-1,2j-1}(j^n2)$ is separable for the
mode decomposition ${\cal A}=\{j,j-1\}$ and 
${\cal B}=\{-j,\cdots ,j\}\setminus\{j,j-1\}$.}.
As a consequence, 
the four modes $m\in\{\pm j,  \pm(j-1)\}$ have vanishing one mode entropy. 
Similar arguments carry over to the modes  $m\in\{\pm j,  \pm(j-2)\}$  in the state, $\Psi_{2j-1,2j-2}(j^n2)$.

\section{Numerical results}\label{numres}

The concept of seniority  has proven  useful for semi-magic
nuclei, where only one type of nucleon is active, and the seniority (i.e., the pseudospin $S$) 
turns out to be conserved with good accuracy. 
 The seniority scheme of the $0f_{7/2}$ and $0g_{9/2}$ subshells, in particular, can be successfully applied to  calcium isotopes \cite{suh,Talmi} and $N=50$ isotones \cite{Talmi}, where
the main prediction of the seniority model, namely that  excitation energies are approximately independent of the particle number
is  fulfilled.
As we now show, entanglement measures detect delicate structures in these correlated quantum states, 
which we can access  through  CI and DMRG calculations, and compare with the predictions of the  seniority  (SEN) model.

Numerical computations were carried out using  the  BIGSTICK code~\cite{joh13} 
and the nuclear shell module of the Budapest DMRG code~\cite{dmrgcode}. 
The BIGSTICK code determines the reduced matrix elements
\begin{equation}
\frac{1}{\sqrt{2K+1}}\langle \Psi_{J}\vert\vert \left[c_{a}^\dagger\otimes\tilde c_{a'}\right]^{(K)}\vert\vert \Psi_{J}\rangle,
\end{equation}
from which we can construct the 1B-RDM and from those the mode entropies using Eq.~(\ref{calcv2a}). 
The DMRG code provides the one-body  density matrix and the one-mode entropies directly.

\begin{figure}
\includegraphics[width=0.85\columnwidth]{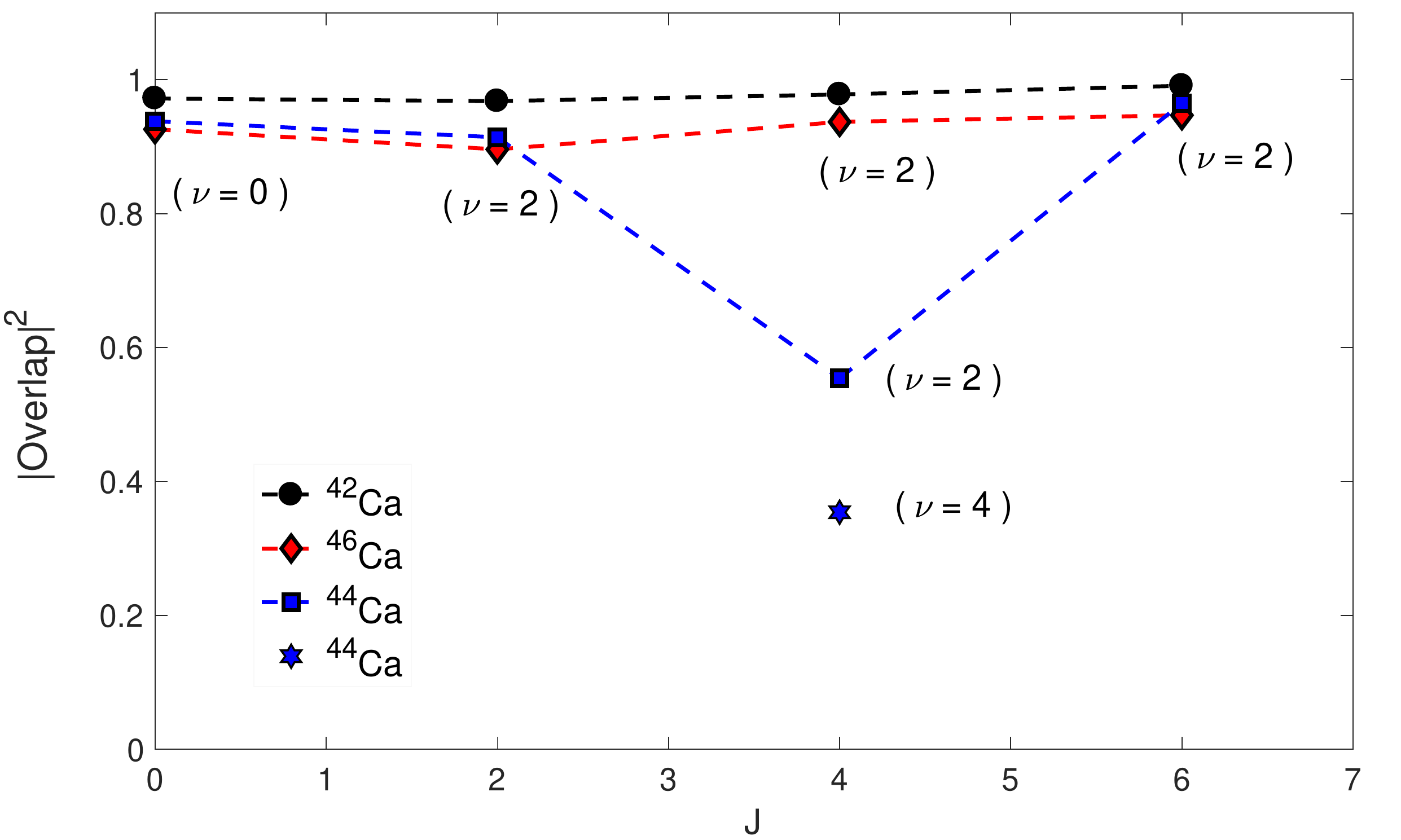}
\caption{\label{oveca}
Squared overlap of the CI  and SEN model wave functions, 
as a function of the total angular momentum for  $^{42}{\rm Ca}$, $^{44}{\rm Ca}$, 
and $^{46}{\rm Ca}$ nuclei. Full CI calculations were performed with the GXPF1
interaction. The   ground  states and the yrast states are very well captured by  $\nu=0$
seniority  and $\nu=2$  seniority SEN states, respectively,  excepting 
the $J=4$ state of $^{44}{\rm Ca}$, where a mixing with seniority $\nu=4$ excitations 
is apparent.}
\centering
\end{figure}

\subsection{Calcium isotopes}

Seniority can be viewed as a quantum number that differentiates between states with the same total angular momentum in the configuration $j^n$. The simplest example is the configuration  $(0f_{7/2})^4$, where there are two $J=2$ and two $J=4$ states with different seniorities ($\nu=2,4$).
According to Ref.~\cite{cap12}, the ground and $J=2$  yrast states  of the nuclei 
$^{42}{\rm Ca}$, $^{44}{\rm Ca}$, $^{46}{\rm Ca}$ are almost pure $(0f_{7/2})^n$ configurations, and the occupations of the other
 $sp$ orbitals are negligible.  In this subsection we  study the mode-entanglement properties of these light calcium isotopes. 
 We always focus on states with maximal $J_z=J$.
  
  We first determined the  occupations 
$\sum_m \langle \Psi_{JJ}\vert c_{am}^\dagger c_{am}\vert \Psi_{JJ}\rangle$
 of $^{44}{\rm Ca}$ for the yrast  states $J=0,2$, and 4 by using the interaction GXPF1 \cite{homa} within a  full CI approach.
For the orbitals  $0f_{7/2}$,  $1p_{3/2}$, $0f_{5/2}$, and  $1p_{1/2}$ we obtain the 
ground state occupation numbers  (3.89, 0.06, 0.05, 0.01), while the occupation numbers  of the $J=2$ and 
$J=4$ states read  (3.89, 0.08, 0.02,0.006) and  (3.89, 0.07, 0.04, 0.005), respectively. Clearly, the four valence shell 
neutrons reside almost exclusively on the $0f_{7/2}$ shell, and occupy the other three shells with very small probabilities. 
Interestingly, the average occupation numbers take similar values for the ground state and the $J=2$ and $J=4$ yrast states.

More detailed information can be gained regarding correlations through the one-mode entropies of the sp orbitals
or from the overlap of the CI wave functions with the SEN model states. 
Fig. \ref{oveca} shows that all many-body eigenstates  are essentially  $\nu=0$ or $\nu=2$ states, with the sole exception
 of the $J=4$ state  of $^{44}{\rm Ca}$. In the  case of $^{44}{\rm Ca}$, it is known that there 
 is strong seniority mixing \cite{Talmi}. We can determine the amplitudes of the seniority $\nu=2$ and $\nu=4$ components  
 by maximizing the overlap between the CI and the mixed seniority  state. 
 By mixing  seniority $\nu=2$ and $\nu=4$ states appropriately, we can increase the overlap   
with the $J=4$  $^{44}{\rm Ca}$ state   to $0.953$. 

\begin{figure}
\includegraphics[width=\columnwidth]{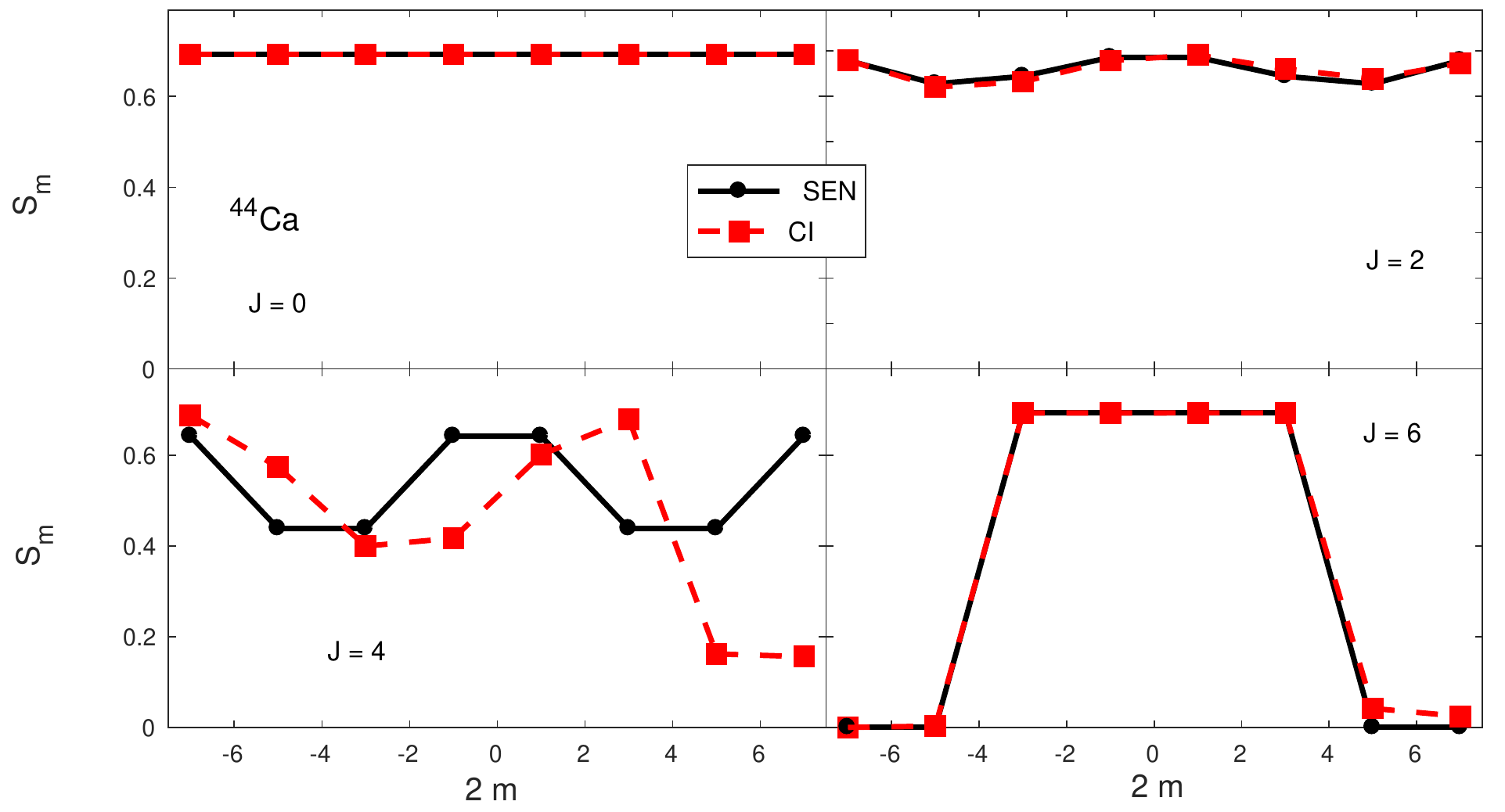}
\caption{\label{omeca44}
Mode entropies of the sp orbital $0f_{7/2}$ for  $^{44}{\rm Ca}$ ground and yrast states.
The results of CI and DMRG calculations  performed with the GXPF1 interaction are compared with 
predictions of  the seniority model (ground state $\nu=0$, yrast states $\nu=2$) . }
\centering
\end{figure}

The precise seniority content of the states and  seniority mixing can be further verified 
by investigating the mode entropies, displayed in Fig.\ref{omeca44} for $^{44}{\rm Ca}$.
Clearly, comparison of the CI /DMRG and SEN  one-mode entropies confirms that the ground and 
$J=2,6$ states of $^{44}{\rm Ca}$ are seniority-zero and seniority-two states, respectively.
As discussed in Section~\ref{anres}, the mode entropies are supposed to reach 
 their maximal value $\ln 2 \approx 0,6931$ in the seniority-zero $J=0$ state of a half-filled  
shell. This prediction is indeed very well satisfied according our full CI / DMRG calculations
in the ground state of $^{44}{\rm Ca}$. Mode entropies are, however, reduced by pair breaking. In particular, in the 
$J=6$  yrast state, the pair-broken   neutrons occupy the $m=7/2$ and $m=5/2$ states, and the mode entropy 
of    the states $m=\pm 7/2$ and $m=\pm 5/2$ is indeed close to zero.

The CI / DMRG one-mode entropies of the $J=4$ state of $^{44}{\rm Ca}$ can not be reproduced with a simple 
seniority $\nu=2$ state and, similar to the $J=4$ state of $^{94}$Ru, discussed later, a mixing of the $\nu=2$ and $\nu=4$
seniority states is needed to reproduce the observed pattern of ${\cal S}_m$.
We mention that  no seniority $\nu=4$ states exist for $^{42}{\rm Ca}$ and $^{46}{\rm Ca}$, 
where the  CI / DMRG results agree well with the SEN model predictions with  $\nu=0$ and $\nu=2$ states only. 

\begin{figure}
\includegraphics[width=0.9\columnwidth]{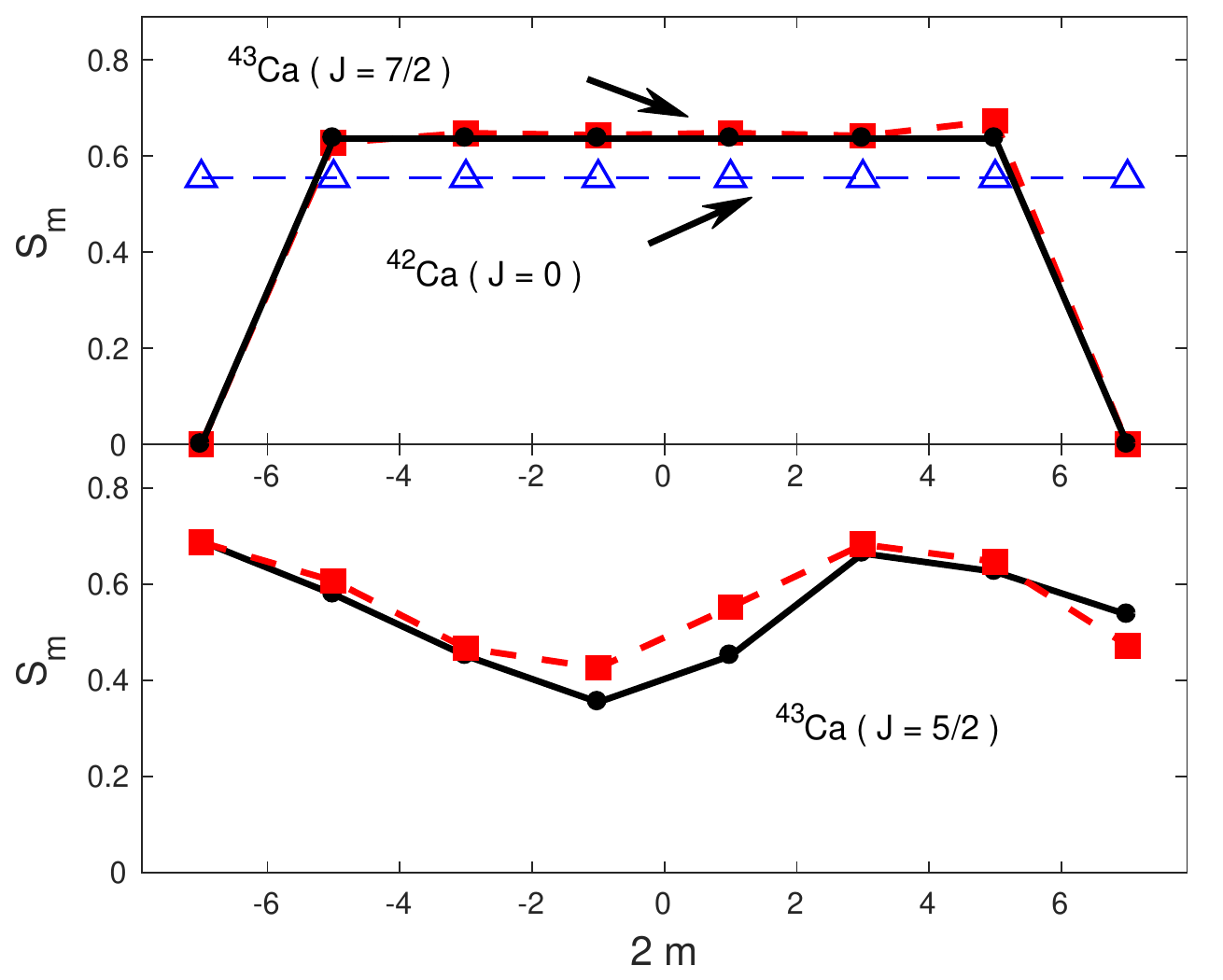}
\caption{\label{caodd}
Mode entropies of the sp orbital $0f_{7/2}$ in the  $J=7/2$  ground state of $^{43}{\rm Ca}$, 
and in its $J=5/2$ yrast state, as determined by CI / DMRG calculations (red squares) and 
predicted by the SEN model (black continuous lines). As a reference, we also display the 
ground state entropies of  $^{42}{\rm Ca}$ (empty triangles).}
\centering
\end{figure}

The seniority model predictions also agree well with the  CI / DMRG results for
 odd calcium nuclei.  To demonstrate this, let us
   consider the ground ($\nu=1$) and first excited states ($\nu=3$) of $^{43}{\rm Ca}$.  
 Increasing the neutron number from 22 to 23  modifies the ground state 
 one-mode entropies as the seniority model 
 predicts (see Fig.~\ref{caodd}). The ground state of $^{43}{\rm Ca}$ has angular momentum $J=7/2$.
 According to Section~\ref{anres}, in the $M=7/2$ seniority $\nu=1$ state, only the modes  $m=-7/2$ and  $m=7/2$ 
have ${\cal S}_m\approx 0$, while the occupation and entropy of the other modes slightly increases. 
The one mode entropies of the excited $J^\pi=({5}/{2})^-$ state are also  well captured by the 
$\nu=3$ SEN model state.  From  Figs.~\ref{omeca44} and \ref{caodd} we thus conclude that 
mode entropy can be used as a sensitive tool to characterize  the seniority structure of 
the ground state and low lying excitations. 

\begin{figure}[b!]
\includegraphics[width=\columnwidth]{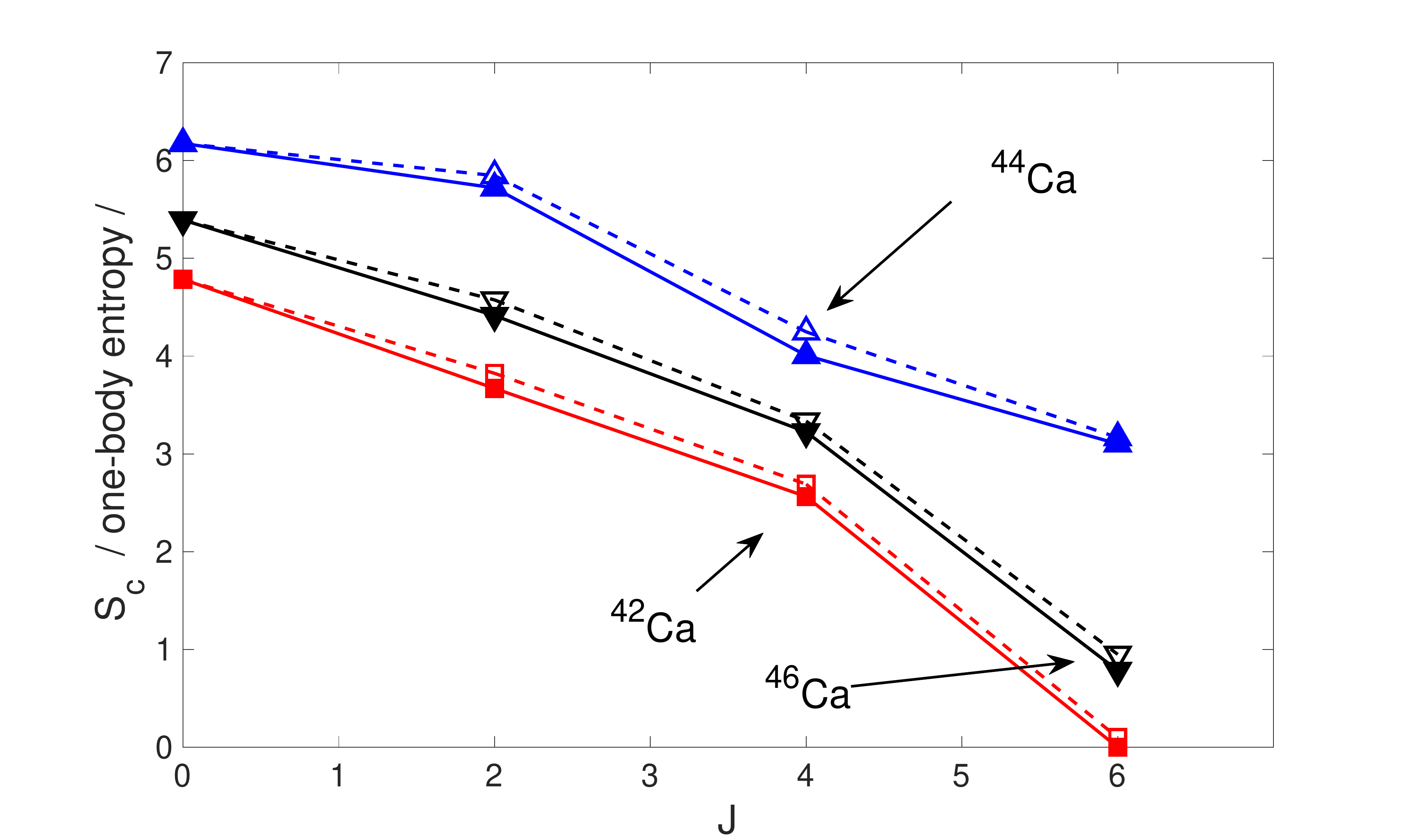}
\caption{\label{totcor}
Total correlations (dashed lines, empty symbols)  and  one-mode entropies (continuous lines, filled symbols)
as a function of  total angular momentum 
for the nuclei $^{42}{\rm Ca}$, $^{44}{\rm Ca}$, and $^{46}{\rm Ca}$. Full CI / DMRG 
calculations were carried out with  the GXPF1 interaction. }
\centering
\end{figure}

{We close this subsection by presenting  
 the total correlations and the basis independent  one-body  entropies
 for the ground and  yrast states of   $^{42}{\rm Ca}$, $^{44}{\rm Ca}$, and  $^{46}{\rm Ca}$ 
 in Fig.~\ref{totcor}. 
 The one-body entropy is obtained after diagonalizing the 1B-RDM, and using the  natural orbitals as a basis.
The total correlation computed using the original shells provides a very good estimate for ${\cal S}_\textrm{1B}$, indicating 
that mixing with other orbitals is rather small. This is only slightly lowered if we use the 
basis of natural orbitals, where the total correlation ${\cal S}_c$ coincides with ${\cal S}_\textrm{1B}$.

 For a given isotope, the ground state has the largest one-body entropy. Excited states 
 contain broken Cooper-pairs, and are less and less entangled with increasing $J$.  
 This latter trend follows from the general entanglement structure of spin states: two coupled 
 spins tend to be most entangled in the smallest, 'antiferromagnetic' spin state, while they are 
 completely unentangled in the largest, 'ferromagnetic' spin state.
   In the $J=0$ ground states,  ${\cal S}_\textrm{1B}$ and ${\cal S}_c$
  agree with each other up to three-four digits. Also, for a fixed $J$, the one-body entropy is maximal 
  for a half-filled shell, i.e. for $^{44}{\rm Ca}$, as predicted by the seniority model. 
  The one-body entropies of $^{42}{\rm Ca}$ and $^{46}{\rm Ca}$ follow  very 
  similar lines, but they are not identical, as particle-hole symmetry would imply. 
  This deviation from the prediction of the seniority model  signals again
  that particle-hole symmetry -- a characteristic property of the single shell SEN model -- 
  is just approximate.}

\vskip0.5cm

\subsection{Entanglement in $^{\rm 94}{\rm Ru}$ nucleus}

As a next example, we  consider  the nucleus $^{\rm 94}{\rm Ru}$ among the $N=50$ isotones.
In case of $N=50$ isotones,  coupling within the $0g_{9/2}$ proton subshell is expected to dominate, 
but contribution from nearby orbitals may also play a role.
In a $0g_{9/2}$ shell-based  seniority model,   
 the seniority quantum number is not enough to uniquely distinguish between states with identical total angular momentum.
In particular, seniority $\nu=4$ states with angular momenta and parity $J^\pi = 4^+$ and $6^+$ are not uniquely defined 
since they both span  two-dimensional subspaces~\cite{ros03}.
As noticed in \cite{esc06,isa08}, there are special seniority $\nu=4$   states with quantum numbers $J^\pi=4^+$ and  $6^+$, 
which have  'good seniority' for any interaction, i.e. states, which  are eigenvectors of both seniority conserving and seniority mixing 
Hamiltonians, when restricted to the  $0g_{9/2}$ shell. These states are called solvable
 \cite{isa14} or $\alpha$ states \cite{qi17}, and 
they do not mix {directly (i.e., in first order)} with other 
seniority $\nu=2$ or seniority $\nu=4$ states of the configuration $(9/2)^4$.

In the CI~/~DMRG description, we used the $0f_{5/2}, 1p_{3/2}, 1p_{1/2}$, and $0g_{9/2}$  sp orbitals 
to span the shell model's active space, and a $^{56}{\rm Ni}$ nucleus as a core.
The one mode entropies  were computed for the ground state, $J^\pi=0^+$, and the yrast states $J^\pi=2^+$, $4^+$, $6^+$, and $8^+$.
For the nucleon-nucleon interaction we  used the so-called jun45 force~\cite{hon09}, and the lisp interaction~\cite{lis04}. 

\begin{figure}[t!]
\includegraphics[width=0.9\columnwidth]{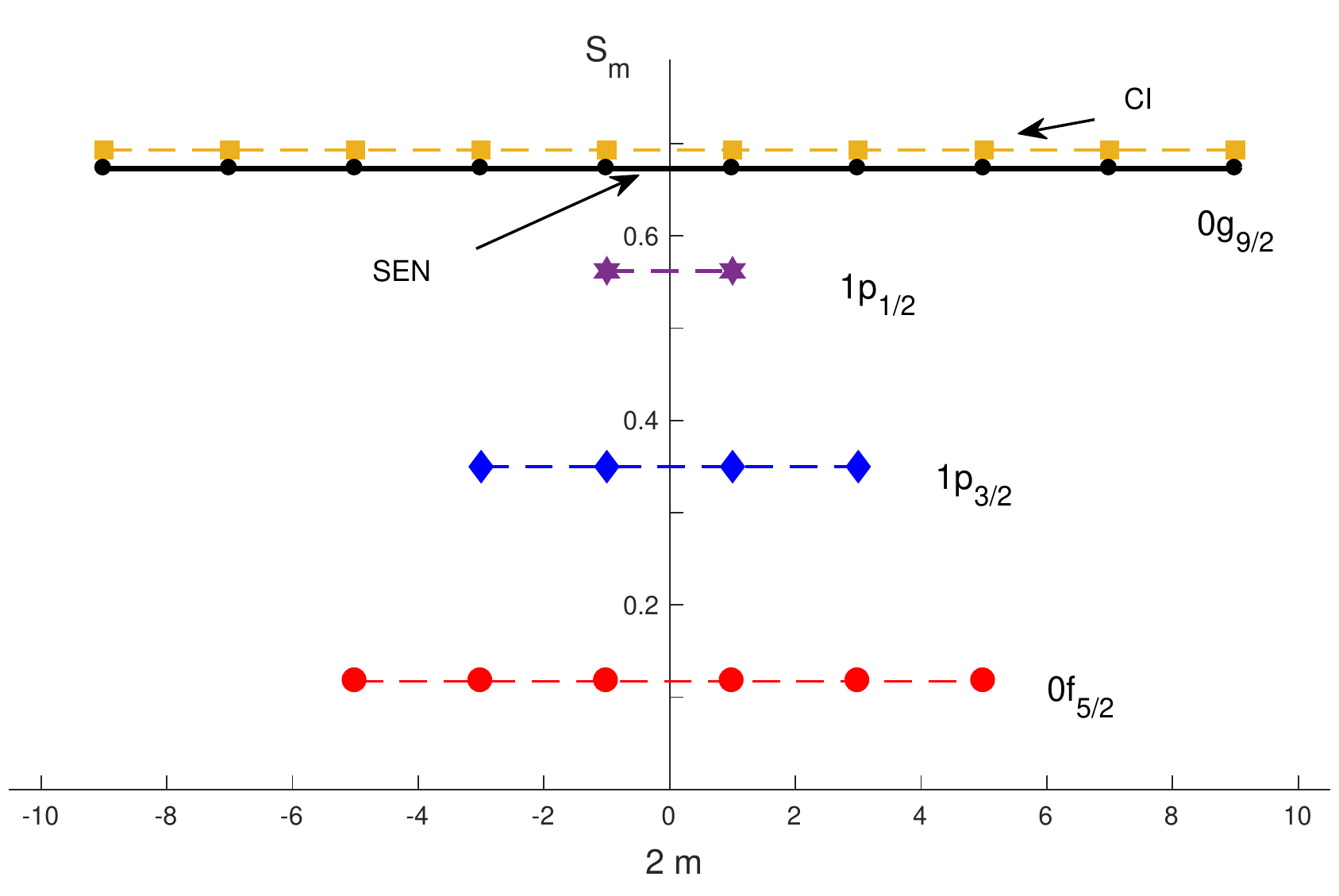}
 \caption{\label{fig:gs}
 Ground state mode entropies of  $^{94}\textrm{Ru}$ sp orbitals obtained  by 
  CI / DMRG   { with the jun45 interaction} (symbols, dashed lines),  
compared with the SEN model prediction (black continuous line).
} 
\centering
\end{figure}

The CI  wave functions contain sixteen protons, whereas the single $j=9/2$ shell SEN wave function has only four.
To compare these two wave functions, we extend the SEN model by simply building on the top of closed $1p_{1/2}^2$, $1p_{3/2}^4$, and 
$0f_{5/2}^6$ shells as 
\begin{equation}\label{cisen}
\hat\Phi_{g_{9/2}} \vert p_{1/2}^2\,p_{3/2}^4\, f_{5/2}^6 \rangle \;,
\end{equation}
with the operator $\hat\Phi_{g_{9/2}} $ creating a four-particle  state $\Psi_{g_{9/2}}$ within the $0g_{9/2}$ shell. 
We refer to seniority states  of this form  as CI-SEN states  or "seniority-like" shell model configurations
in the following. Clearly, the one mode  entropies associated with  the filled $1p_{3/2}$, $1p_{1/2}$, and $0f_{5/2}$ shells vanish, 
while  the mode entropies of the  $0g_{9/2}$ orbitals  in a CI-SEN state are identical to  those of the corresponding 
four-particle $j=9/2$ SEN state.

\begin{table}[b]
\begin{center}
\begin{tabular}{|c|c|c|c|c|c|}
\hline
$J^\pi$&model&jun45&lisp\\
\hline
\multirow{2}{*}{$0^+$}
&$\Psi_{00}(j^40)$&0.517&0.600\\
&$\Psi_{00}(j^44)$&$2.1\times 10^{-4}$&$7.3 \times 10^{-5}$\\
\hline
\multirow{2}{*}{$2^+$}
&$\Psi_{22}(j^42)$&0.633&0.696\\
&$\Psi_{22}(j^44)$&$9.4\times 10^{-5}$&$3.2\times 10^{-5}$\\
\hline
\multirow{3}{*}{$4^+$}
&$\Psi_{44}(j^42)$&0.416&0.724\\
&$\Psi{44}(j^44 \,\alpha)$&0.277&$3.1\times 10^{-4}$\\
&$\Psi{44}(j^44\, \beta)$&$5.5\times 10^{-4}$&$4.1\times 10^{-4}$\\
&mix&0.692&0.724\\
\hline
\multirow{2}{*}{$6^+$}
&$\Psi_{66}(j^42)$&0.646&0.726\\
&$\Psi_{66}(j^44\, \alpha)$&$2.2\times 10^{-4}$&$2.3\times 10^{-5}$\\
&$\Psi_{66}(j^44 \,\beta)$&$6.1\times 10^{-4}$&$2.8\times 10^{-4}$\\
\hline
\multirow{2}{*}{$8^+$}
&$\Psi_{88}(j^42)$&0.657&0.727\\
&$\Psi_{88}(j^44)$&$4.0\times 10^{-5}$&$1.6\times 10^{-5}$\\
\hline
\end{tabular}
\end{center}
\caption{\label{overlap}
\label{table:overlaps}
{Square of the modulus of the overlap of     different CI-SEN states with the CI wave functions, 
 computed  using the interactions jun45 \cite{hon09} and lisp \cite{lis04}.}
} 
\end{table}

To quantify the seniority content of a CI   wave function, we   calculate the square of the modulus of the overlap 
of the CI wave function and a given CI-SEN state, i.e., the overlap probabilities. Results are listed  in Table~\ref{table:overlaps}. 
For  $J^\pi=4^+$ and  $J^\pi=6^+$, Table~\ref{overlap} includes overlaps with the aforementioned 
solvable or $\alpha$ states  and with the $\beta$ states. The  $\Psi_{JM}(j^44\,\beta)$  states are four-particle seniority-four states with $J=4,6$ such that they are orthogonal to the corresponding solvable  states.
In all cases,  seniority-four states are slightly mixed in, and the optimal overlap is reached with a state 
\begin{equation}\label{mixmssen}
{|\Psi_\textrm{mixed}\rangle = \sum_{\nu,\eta} v_{\nu\eta}\,  \hat\Psi_{JM}(j^4\nu\eta)\,\vert p_{1/2}^2\,p_{3/2}^4\,f_{5/2}^6 \rangle\,.}
\end{equation}
In most cases, similar to the Ca isotopes, 
the seniority $\nu=0$ and $\nu=2$ states dominate, and the contribution of seniority $\nu=4$ states is negligible. 
 {An} exception is the state with  $J^\pi=4^+$ computed by using the jun45 interaction, where 
the contribution of  the seniority-four $\alpha$ state is significant.
Since the overlap with the $\beta$ state turns out to be very small, mixing calculations for states of the form  (\ref{mixmssen}),
presented in  Table~\ref{table:overlaps}, have been restricted to  solvable ($\alpha$) states only.

\begin{figure}[b]
\includegraphics[width=0.9\columnwidth]{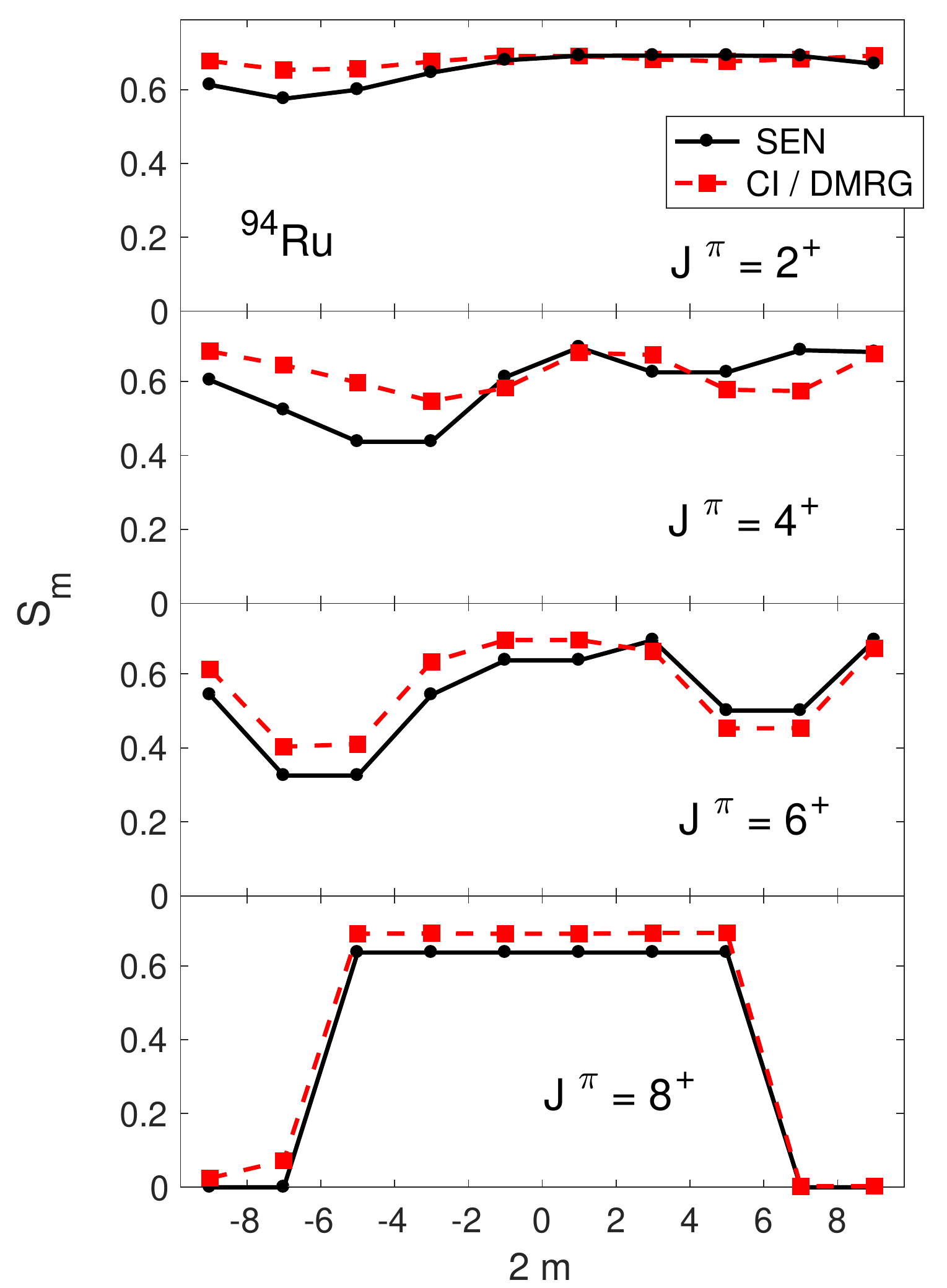}
\caption{\label{fig:excited}
Mode entropies   of the sp orbital $0g_{\frac{9}{2}}$ in the yrast excited states of  $^{94}{\rm Ru}$, constructed 
using { full CI}  and DMRG { with the jun45 interaction (red squares)}, and compared with seniority $\nu=2$ states of the 
SEN model {(black continuous lines)}. 
} 
\centering
\end{figure}

In  the $4^+$ state, the jun45 interaction generates  strong seniority mixing, and  the square of the modulus of the overlap
 of the CI wave function and the seniority mixed CI-SEN state is maximal for 
 the amplitudes  $\vert v_{2}\vert^2=0.593$, $\vert v_{4,\alpha}\vert^2=0.407$.
  This observation agrees with the results of  Ref.~\cite{qi17}, where it was shown that these properties of the jun45 
 wave function can explain the observed $E2$ transition probabilities of $^{94}{\rm Ru}$. 

We emphasize that  the mixing of the states $\hat\Psi_{JM}(j^42)\vert 0\rangle$ and $\hat\Psi_{JM}(j^44)\vert 0\rangle$ is not possible if
interactions are strictly restricted to the  $j=9/2$ shell. Interactions can, however, mix the seniority-two 
and seniority-four components due to the presence  of other shells~\cite{qi17}.

As we have seen, the CI wave functions are reasonably well described in terms of mixed CI-SEN  states. 
Particle number fluctuations on the $1p_{3/2}$, $1p_{1/2}$, and $0f_{5/2}$ shells are, however, non-negligible, and these 
shells are  thus not completely filled.  This is clearly shown by the ground state  mode entropies, 
 displayed in Fig.~\ref{fig:gs}. Since the ground state has quantum numbers $J^\pi = 0^+$, 
 mode entropies are independent of  the magnetic quantum number $m$ in this case within any shell.  
 The   CI / DMRG mode entropies in the  $0g_{9/2}$ shell are 
   almost identical to those of the CI-SEN model. 
However,   the  mode entropies of the $1p_{3/2}$, $1p_{1/2}$, and $0f_{5/2}$ shells are 
 relatively large in the CI /DMRG calculations,  implying that these shells have substantial 
 proton number  fluctuations, induced 
 by  nucleon-nucleon interactions. Indeed,  the mode entropies are directly related to the occupation 
 probabilities of these modes,  numerically computed as  
 $P_{m,3/2} = 0.8884$,  $P_{m,5/2} = 0.9749$, and  $P_{m,1/2} = 0.7502$
by using the jun45 interaction.

Mode entropies of  the yrast states obtained using the jun45 interaction 
are shown in Fig.~\ref{fig:excited}, and compared with the 
predictions of  the SEN model using seniority $\nu=2$ states only (no seniority mixing). 
We display only mode entropies for  the sp state, $0g_{9/2}$.
The two models give similar patterns for the one-mode entropies, and the quantitative agreement is also satisfactory except for 
 the  $4^+$ state,  where the non-mixing  SEN model has a somewhat larger deviation with respect to  
 CI and DMRG computations.

As shown in  Fig.~\ref{fig3},  neither the seniority-two SEN model,  nor the seniority-four solvable state can explain 
the shape of the one mode entropies of the CI {and DMRG} computations. One 
 must use the seniority mixed CI-SEN wave function to obtain  a better agreement, and indeed, 
the optimal  mixing amplitude $\vert v_{2}\vert^2=0.723$, $\vert v_{4,\alpha}\vert^2=0.277$
produces a  quite satisfactory agreement. 
The remaining relatively small discrepancies between the two calculations can be attributed
 to the fact that the full CI / DMRG wave states contain,  of course, excitations and configurations
 beyond the  CI-SEN components.

\begin{figure}
\includegraphics[width=\columnwidth]{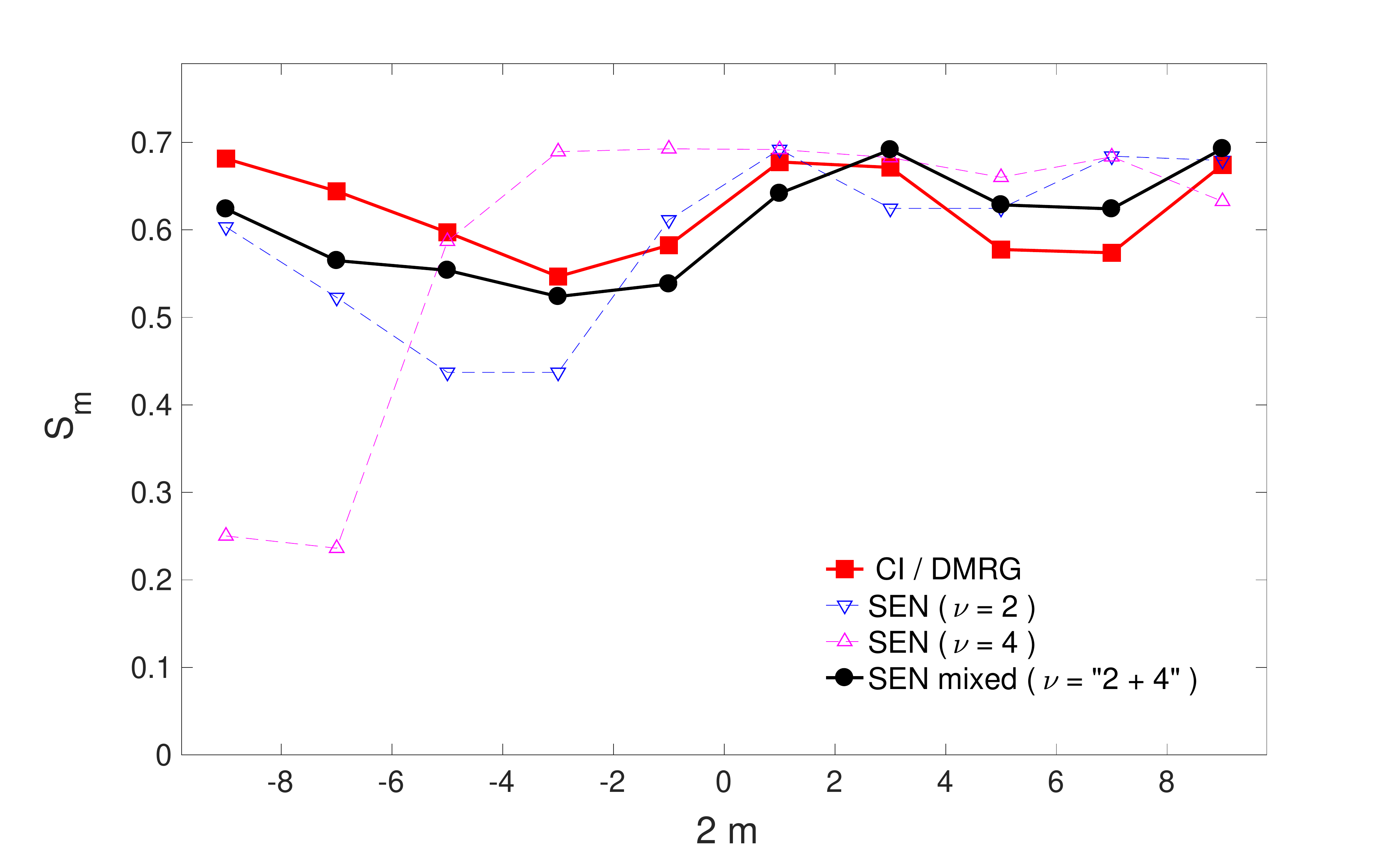}
\caption{\label{fig3}
Mode entropies of the $^{94}$Ru sp orbital $0g_{{9/2}}$, calculated for the yrast state
 $J^\pi = 4^+$  using {full} CI/DMRG computations { with the jun45 interaction}, 
and compared with the $0g_{{9/2}}$ shell SEN model. Mixing with 
the so-called $\alpha$ seniority-four state accounts for the observed structures, and 
is needed to yield satisfactory agreement.}
\begin{center}
\end{center}
\end{figure}

\section{Summary}\label{sum}

 In this work,  we analyzed the entanglement structure of the open shells of 
 certain semi-magic  nuclei, and compared the observed structures with the 
 predictions of a single-$j$ shell  seniority (SEN) model. 
 
  We first derived analytical expressions for the one-body reduced density matrix  within the 
  SEN  model for states with seniority zero, one, two, and three. We determined 
the particle number dependence  of the  one-body reduced density matrix for arbitrary seniority, and
we have shown that,  within the $j^n$ configuration space, 
wave functions of angular momentum $J=0$  have maximal one-body entanglement entropy, irrespective of 
the seniority.  Breaking Cooper-pairs, and aligning the angular momenta of the pair-broken 
nuclei  reduces  the one-body entropy, and the one-body entropy is found to decrease with 
increasing $J$ for a given nucleus. 
 
The seniority model  predicts peculiar properties for  half-filled shells,
also manifest in entanglement measures. 
For seniority-zero and seniority one states, the one-body entanglement entropy 
is maximal  for  a half-filled shell, $n = (2j+1)/2$. Also, in the seniority model, the 
entropy displays particle-hole symmetry: $n$-particle and $n$-hole ($2j+1-n$) particle states
are predicted to have identical one-body entanglement entropies.

We carried out full CI  and  numerical density matrix renormalization group  (DMRG) calculations
for $^{42}{\rm Ca}$~,$\dots$, $^{46}{\rm Ca}$,  isotopes and for  $^{94}{\rm Ru}$, and  
compared the numerical results with the predictions of the corresponding  $(0f_{7/2})$ and
 $(0g_{9/2})$ shell seniority models. 
Mode entropies show an overall good agreement  for the ground   and  yrast states 
with a few exceptions, where  clear signatures of 
seniority mixing are observed. 

We first verified the predictions of the seniority model for Ca isotopes. 
For all even isotopes ($^{42}{\rm Ca}$, $^{44}{\rm Ca}$, and $^{46}{\rm Ca}$), the one-body entanglement is 
maximal in the $J=0$ ground state,  takes a value very close to the one predicted by the seniority model,
and decreases with increasing $J$. As predicted by the seniority model, for a given $J$, one-body entanglement 
is maximal for a half-filled shell ($^{44}{\rm Ca}$), and an approximate particle-hole symmetry 
is observed between  the $^{42}{\rm Ca}$ and $^{46}{\rm Ca}$ isotopes. 
The breaking of particle-hole symmetry can be attributed 
to neutron number fluctuations  on deeper shells, which have fractional occupations
and, correspondingly, exhibit   sizable one-body entropies.

The full CI / DMRG wave functions of the ground  and yrast states  have  
large overlaps with seniority-like $(0f_{7/2})^4$  configurations, 
with the  dominant components having seniority $\nu=0$ (for $J=0^+$) and $\nu=2$ (for $J=2$ and $J=6$). 
In case of the $J=4$, yrast state of  $^{44}{\rm Ca}$,  however, strong  mixing is observed 
with the $\nu=2$ and $\nu=4$ states.  This mixing, generated by neutron number 
fluctuations on other shells,  turns out to be  essential to explain the fine structures of mode entanglement. 

A similar mixing pattern is observed in $^{94}{\rm Ru}$, where  the full CI wave functions of the ground 
and yrast states are found to have  large overlaps with seniority-like $(0g_{9/2})^4$  configurations.
The  dominant components have also seniority $\nu=0$ (for $J^\pi=0^+$) or $\nu=2$ (for $J^\pi=2^+$ and $J^\pi=6^+$).  
Similar to $^{44}{\rm Ca}$,  however, the $4^+$ yrast state of $^{94}{\rm Ru}$ displays strong seniority mixing 
with the so-called solvable seniority $\nu=4$ state (or $\alpha$ state). 
Similar to the neutron shells of Ca, the $1p_{1/2}$, $1p_{3/2}$, and 
$0f_{5/2}$ proton shells exhibit sizable one-body entropies, and display corresponding fractional occupations. 
Our findings are in line with earlier observations\cite{qi17} that mixing 
 with the  solvable seniority $\nu=4$ state may be significant due to the presence of 
other, non seniority-like configurations,  and is essential  to explain the BE(2) 
transition probabilities of  $^{94}{\rm Ru}$. 

Mode and one-body entropies are thus extremely useful tools 
to investigate the structure of quantum correlations in nuclei. Here we restricted our discussions 
to   simple semi-magic nuclei, where the seniority model provides an appropriate
analytical framework and reference point. Extending our approach to 
 study quantum fluctuations and quantum correlations in generic, open shell 
 nuclei represent exciting perspectives for future research.  

This work has been supported by the National Research, Development and Innovation Fund of Hungary (NKFIH) through Grants
Nos. K128729, K120569, K134983, SNN139581, 
by the Hungarian Quantum Technology National Excellence Program (Project
No. 2017-1.2.1-NKP-2017-00001) and by the Quantum Information National Laboratory of Hungary. {\"O}.L. also acknowledges financial 
support from the Hans Fischer Senior Fellowship programme funded by the Technical University of Munich -- Institute 
for Advanced Study. The development of DMRG libraries has been supported by the Center for Scalable and Predictive methods for 
Excitation and Correlated phenomena (SPEC), which is funded as part of the Computational Chemical Sciences Program by the U.S. 
Department of Energy (DOE), Office of Science, Office of Basic Energy Sciences, Division of Chemical Sciences, Geosciences, 
and Biosciences at Pacific Northwest National Laboratory.

\appendix
\section{Matrix elements of the 1B-RDM}
\label{appendix}
Here we analytically calculate the matrix elements of the quasi-spin tensor operators $R_{m,-m';00}$ and $R_{m,-m';10}$, which has to be used in the reduction formula (\ref{freduct})
for states with $\nu=0,1,2$.
\begin{widetext}
Equation (\ref{freduct}) shows that we have to calculate matrix elements of the following operators:
\begin{equation}\label{op1}
R^0_{0;m,-m'}=\frac{1}{\sqrt{2}}\left[(-1)^{j+m'}a_{j,m}^\dagger a_{j,m'}-(-1)^{j-m}\left(\delta_{m,m'}-a^\dagger_{j,-m'}a_{j,-m}\right)
\right]
\end{equation}
and 
\begin{equation}\label{op2}
R^1_{0;m,-m'}=\frac{1}{\sqrt{2}}\left[(-1)^{j+m'}a_{j,m}^\dagger a_{j,m'}+(-1)^{j-m}\left(\delta_{m,m'}-a^\dagger_{j,-m'}a_{j,-m}\right)
\right].
\end{equation}
\end{widetext}

In the case of $\nu=0$ according to the reduction formula (\ref{freduct}), we need only the following matrix elements
\begin{equation}
\langle 0\vert R^0_{0;m,-m'}\vert 0\rangle=-\delta_{m,m'}\frac{ (-1)^{j-m}}{\sqrt{2}}
\end{equation} 
and
\begin{equation}
\langle 0\vert R^1_{0;m,-m'}\vert 0\rangle=\delta_{m,m'}\frac{ (-1)^{j-m}}{\sqrt{2}}
\end{equation}
Substituting the last two expressions  into (\ref{freduct}), we get (\ref{nu0om}). 

When the seniority is one, the corresponding wave function that enters into the reduction formula (\ref{freduct})  is $\Psi_{jM}(j^11)=a^\dagger_{jM}\vert 0\rangle$. 
Simple calculation gives
\begin{eqnarray}
&\langle \Psi_{jM}(j^11)\vert R^0_{0;m,-m'}\vert \Psi_{jM}(j^11)\rangle
=\delta_{m,m'}\frac{(-1)^{j-m}}{\sqrt{2}}\nonumber\\
&(-1+\delta_{-m,M}-
\delta_{m,M})
\end{eqnarray} 
and
\begin{eqnarray}
&\langle \Psi_{jM}(j^11)\vert R^1_{0;m,-m'}\vert\Psi_{jM}(j^11)\rangle=\delta_{m,m'}\frac{(-1)^{j-m}}{\sqrt{2}}\nonumber\\
&(1-\delta_{-m,M}-
\delta_{m,M})
\end{eqnarray} 
Using these last two expressions and (\ref{freduct}), we can derive (\ref{nu1om}).
 
In order to calculate the 1B-RDM in the case of seniority equals to two, we have to determine the matrix elements of the operators $R_{m,-m';00}$ and $R_{m,-m';10}$ 
between two-particle wave functions. 
The normalised two-body wave function with total angular momentum $J$ and projection $M$ is
\begin{equation}\label{wf2}
\Psi_{JM}(j^22)=\frac{1}{\sqrt{2}}\sum_{m} C_{ j,m,j,M-m}^{J,M} a_m^\dagger a^\dagger_{M-m}\vert 0 \rangle ,
\end{equation}
where $J=2,4,\ldots,2j-1$.
The building blocks are the matrix elements
\begin{equation}\label{wf2res}
\langle\Psi_{JM}(j^22)\vert a^\dagger_{m} a_{m'} \vert\Psi_{JM}(j^22)\rangle=\delta_{m,m'}2
\left(C_{ j,m,j,M-m}^{ J,M}\right)^2
\end{equation}
which can be obtained with the use of the Wick theorem.
\begin{widetext}
Finally from (\ref{op1}), (\ref{op2}) and (\ref{wf2res}) we derive 
\begin{eqnarray}\label{aaa1}
&\langle\Psi_{JM}(j^22)\vert R^0_{0;m,-m'} \vert\Psi_{JM}(j^22)\rangle=\nonumber\\
&\delta_{m,m'}(-1)^{j-m}\frac{1}{\sqrt{2}}\left[-2\left(C_{ j,m,j,M-m}^{J,M}\right)^2
+2\left(C_{ j,-m,j,M+m}^{J,M}\right)^2-1\right]\nonumber\\
\end{eqnarray}
and
\begin{eqnarray}\label{aaa2}
&\langle\Psi_{JM}(j^22)\vert R^1_{0;m,-m'} \vert\Psi_{JM}(j^22)\rangle=\nonumber\\
&\delta_{m,m'}(-1)^{j-m}\frac{1}{\sqrt{2}}\left[-2\left(C_{ j,m,j,M-m}^{ J,M}\right)^2
-2\left(C_{ j,-m,j,M+m}^{J,M}\right)^2+1\right]\nonumber\\
\end{eqnarray}
\end{widetext}
If we use the equation (\ref{aaa1}) and (\ref{aaa2}) the reduction formula (\ref{freduct}) we get  the 1B-RDM in the form (\ref{nu2om}).

In order to calculate the 1B-RDM for seniority-three states first we rewrite the general expression 
(\ref{freduct}) in the form
\begin{eqnarray}\label{freduct2}
&\rho_{m',m}(JM,j^n\nu\rho)=-\delta_{m,m'}\frac{1}{2}
\left[\left(-A_{m,m}^\nu+A^\nu_{-m,-m}-1\right)\right.\nonumber\\
&\left.+\frac{\Omega-n}{\Omega-\nu}\left(-A_{m,m}^\nu-A^\nu_{-m,-m}+1\right)\right],
\end{eqnarray} 
where
\begin{equation}
A_{m,m}^\nu=\langle\Psi_{JM}(j^\nu\nu\rho)\vert c_m^\dagger c_m\vert\Psi_{JM }(j^\nu\nu\rho)\rangle
\end{equation}

A general pure three-particle state can be written in the form \cite{eck02}
\begin{equation}\label{3pstate}
 \Psi = \sum_{ijk} w_{ijk} c^\dagger_i  c^\dagger_j c^\dagger_k \vert 0\rangle,
\end{equation}
where the coefficients $w_{ijk}$ of the superposition are fully antisymmetric.
Using the antysimmetry property of the coefficients $w_{ijk}$ and the commutation relations of the creation operators we can get the following expression
\begin{equation}\label{3pformula}
\langle \Psi \vert c^\dagger_x c_y \vert \Psi \rangle = 18\sum_{jk}w_{xjk}w_{yjk}.
\end{equation}

A seniority-three state can be turned into the form \cite{kan80,che19}
\begin{eqnarray}\label{sen3}
&\Psi_{JM}(j^33 J_2)=\frac{1}{\cal N}
\left(\left[a^\dagger \otimes \left [a^\dagger\otimes a^\dagger \right]^{(J_2)}\right]^{(J)}_M \vert 0\rangle\right.\nonumber\\
&\left.+A_3 \left[ \left [a^\dagger\otimes a^\dagger \right]^{(0)} \otimes a^\dagger \right]^{(J)}_M \vert 0\rangle\right),
\end{eqnarray}

where $J_2$ is even and positive and 
\begin{equation}
{\cal N}=\left[1+2(2J_2+1){
 \begin{Bmatrix} 
   j & j & J_2  \\
   j & J & J_2  \\
   \end{Bmatrix} 
}-\frac{4\delta_{jJ}(2J_2+1)}{(2j+1)(2j-1)}\right]^{\frac{1}{2}},
\end{equation}
\begin{equation}
 A_3=\frac{2\sqrt{2J_2+1}}{2j-1}\delta_{jJ}.
\end{equation}
Now we apply the general expression (\ref{3pformula}) and after a lengthy but straightforward calculation 
get
\begin{widetext}
\begin{eqnarray}\label{sen3formula}
\langle \Psi \vert a^\dagger_m a_{m'} \vert \Psi \rangle &=&\delta_{m,m'} \frac{2}{{\cal N}^2}\bigg \{
  \sum_{a} 
\bigg(C_{j,M-m-a,J_2,m+a}^{J,M}C_{j,m,j,a}^{J_2,m+a} 
+C_{j,m,J_2,M-m}^{J,M}C_{j,a,j,M-m-a}^{J_2,M-m}+C_{j,a,J_2,M-a}^{J,M}C_{j,M-m-a,j,m}^{J_2,M-a}\bigg)^2
 \nonumber\\
 &
 +&\frac{2A_3}{\sqrt{2j+1}}
 \bigg[2(-1)^{j-m}\bigg(C_{j,M,J_2,0}^{J,M}C_{j,m,j,-m}^{J_2,0}+C_{j,m,J_2,M-m}^{J,M}C_{j,-m,j,M}^{J_2,M-m}+
 C_{j,-m,J_2,M+m}^{J,M}C_{j,M,j,m}^{J_2,M+m}\bigg)
 \nonumber\\
 &+&\delta_{mM}\sum_{a}(-1)^{j-a}\bigg(C_{j,-a,J_2,m+a}^{J,M}C_{j,m,j,a}^{J_2,m+a}
 +C_{j,m,J_2,0}^{J,M}C_{j,a,j,-a}^{J_2,0}+
 C_{j,a,J_2,m-a}^{J,M}C_{j,-a,j,m}^{J_2,m-a}\bigg)
  \bigg] 
   \nonumber\\
& +&\frac{A_3^2}{2j+1}\bigg[2+(2j-3)\delta_{mM}-2\delta_{m,-M}\bigg] \bigg \}.
\end{eqnarray}
\end{widetext}

\end{document}